\documentclass[conference, a4paper]{IEEEtran}

\usepackage{array, calc, color, multicol}
\usepackage{algorithm}
\usepackage[noend]{algorithmic} 
\usepackage{multirow}

\usepackage{mathtools}

\usepackage{graphics, epstopdf, graphicx, epsfig, psfrag, epsf, subfigure}
\usepackage{amssymb, amsfonts, amsma th, times}
\usepackage{relsize}
\usepackage{multicol,balance, verbatim}   
\usepackage{textcomp, mathrsfs, stmaryrd, setspace}
\usepackage{amssymb}
\usepackage{wrapfig}
\usepackage{xcolor}
\usepackage{cite}
\usepackage{xspace}

\newcommand{\ie}{{\em i.e., }}

\newtheorem{theorem}{Theorem}
\newtheorem{lemma}[theorem]{Lemma}
\newtheorem{proposition}[theorem]{Proposition}

\newtheorem{example}{Example}

\DeclareGraphicsExtensions{.eps, .pdf, .png, .jpg}

\newcommand{\Nset}{\mathcal{N}}

\newcommand{\Zset}{\mathcal{Z}}

 \newcommand{\SC}{SC$^3$\xspace}

\newcommand{\Co}{ C}

\IEEEoverridecommandlockouts

\begin{document}

\title{Secure Coded Cooperative Computation at the Heterogeneous Edge against Byzantine Attacks}

\author{\IEEEauthorblockN{
Yasaman Keshtkarjahromi\IEEEauthorrefmark{1}, Rawad Bitar\IEEEauthorrefmark{2}, Venkat Dasari\IEEEauthorrefmark{3}, Salim El Rouayheb\IEEEauthorrefmark{2} and Hulya Seferoglu\IEEEauthorrefmark{4}}
\IEEEauthorblockA{\small yasaman.keshtkarjahromi@seagate.com, rawad.bitar@rutgers.edu, venkateswara.r.dasari.civ@mail.mil,\\ salim.elrouayheb@rutgers.edu, hulya@uic.edu}
\IEEEauthorblockA{ \IEEEauthorrefmark{1}Seagate Technology, Storage Research Group,
\IEEEauthorrefmark{2}Rutgers University, New Jersey,\\
\IEEEauthorrefmark{3}US Army Research Lab,
\IEEEauthorrefmark{4}University of Illinois at Chicago}
\thanks{This work was supported in parts by the Army Research Lab (ARL) under Grant W911NF-1820181, National Science Foundation (NSF) under Grants CNS-1801708 and CNS-1801630, and the National Institute of Standards and Technology (NIST) under Grant 70NANB17H188.}
\\\vspace*{-1.2cm}
}

\maketitle

\allowdisplaybreaks

\allowdisplaybreaks

\begin{abstract}
Edge  computing  is  emerging  as  a  new  paradigm  to  allow processing  data  at  the edge  of  the  network,  where  data  is typically  generated  and  collected,  by exploiting multiple devices at the edge collectively. However, offloading tasks to other devices leaves the edge computing applications at the complete mercy of an attacker. One of the attacks, which is also the focus of this work, is Byzantine attacks, where one or more devices  can  corrupt the offloaded tasks. Furthermore, exploiting  the  potential  of  edge  computing  is challenging mainly due to the heterogeneous and time-varying nature of the devices at the edge. In this paper, we develop a  secure coded cooperative computation mechanism (\SC) that  provides both  security  and  computation  efficiency guarantees by gracefully combining homomorphic hash functions and coded cooperative computation. Homomorphic hash functions are used against Byzantine attacks and coded cooperative computation is used to improve computation efficiency when edge resources are heterogeneous and time-varying. 
 
Simulations results show that \SC improves task completion delay significantly.
\end{abstract}

\section{\label{sec:introduction} Introduction}
Edge computing is emerging as a new paradigm to allow processing data at the edge of the network, where data is typically generated and collected. This paradigm advocates offloading tasks from an edge device to other edge/end devices including mobile devices, and/or servers in close proximity. Edge computing can be used in Internet of Things (IoT) applications which connects an exponentially increasing number of devices, including smartphones, wireless sensors, and health monitoring devices at the edge.
Many IoT applications require processing the data collected by these devices through computationally intensive algorithms with stringent reliability, security and latency constraints. In many scenarios, these algorithms cannot be run locally on computationally-limited IoT-devices.

One of the existing solutions to handle computationally-intensive tasks is computation offloading, which advocates offloading tasks to remote servers or to cloud computing platforms. Yet, offloading tasks to remote servers or to the cloud could be a luxury that cannot be afforded by most edge applications, where connectivity to remote servers can be expensive, energy consuming, lost or compromised. In addition, offloading tasks to remote servers may not be efficient in terms of delay, especially when data is generated and collected at the edge. This makes edge computing a promising solution to handle computationally-intensive tasks, where the task is divided into sub-tasks and each sub-task is offloaded to an edge device for computation. 

However, offloading tasks to other devices leaves the edge computing applications at the complete mercy of an attacker. One  of  the  attacks,  which is the focus of this work, is {\em Byzantine attacks}, where one or more devices (workers) can corrupt the offloaded tasks. Furthermore, exploiting the potential of edge computing is challenging mainly due to the heterogeneous and time-varying nature of the devices at the edge. Thus, our goal is to develop a secure, dynamic, and heterogeneity-aware edge computing mechanism that provides both security and computation efficiency guarantees. 

Our key tool is the graceful use of coded cooperative computation and homomorphic hash functions. Coded computation advocates mixing data in computationally-intensive tasks by employing erasure codes and offloading these coded tasks to other devices for computation \cite{KS18,BPR17,li2016unified,dutta2017coded,yang2017computing,halbawi2017improving,yu2017polynomial,DCG16,tandon2017gradient,li2016fundamental,lee2018speeding,karakus2017straggler,aktas2017effective}. The following canonical example demonstrates the effectiveness of coded computation. 

\begin{example} \label{ex:ex1}
Consider the setup where a master device  wishes to offload a task to 3 workers. The master has a large data  matrix $A$  and wants to compute matrix vector product $A\mathbf{x}$. 

The master device divides the matrix $A$ row-wise equally into two smaller matrices $A_1$ and $A_2$, which are then encoded using a $(3,2)$ Maximum Distance Separable (MDS) code\footnote{An $(n,k)$ MDS code divides the master's data into $k$ chunks and encodes it into $n$ chunks ($n>k$) such that any $k$ chunks out of $n$ are sufficient to recover the original data. } to give  $B_1=A_1$, $B_2=A_2$ and $B_3=A_1+A_2$, and sends each to a different worker. Also, the master device sends $\mathbf{x}$ to workers and asks them to compute $B_i\mathbf{x}$, $i \in \{1,2,3\}$. When the master receives the computed values (\ie $B_i\mathbf{x}$) from at least two out of three workers, it can decode its desired task, which is the computation of $A\mathbf{x}$. The power of coded computations is that it makes $B_3=A_1+A_2$ act as a ``joker" redundant task that can replace any one of the other two tasks  if a worker ends up straggling, \ie being slow or unresponsive.
\hfill $\Box$
\end{example}

This example demonstrates the benefit of coding for edge computing. However, the very nature of task offloading to workers makes the computation framework vulnerable to attacks. We focus on Byzantine attacks 
in this work. For example, if workers $1$ and $3$ in Example \ref{ex:ex1} corrupt $B_1\mathbf{x}$ and $B_3\mathbf{x}$, the master can only obtain a wrong value of $A\mathbf{x}$. Thus, it is crucial to develop a secure coded computation mechanism for edge devices against this type of attacks. 

In this paper, we develop a secure coded cooperative computation (\SC) mechanism which uses homomorphic hash functions. Example~\ref{ex:ex2} illustrates the main idea of homomorphic hash functions in coded computation. 

\begin{example} \label{ex:ex2}
Consider the same setup in Example \ref{ex:ex1}, and assume that worker $i$ returns the computed value $\tilde{\mathbf{y}}_i$ to the master device. If worker $i$ is an honest worker,  $\tilde{\mathbf{y}}_i =  B_i\mathbf{x}$ holds. 
The master device checks the integrity of $\tilde{\mathbf{y}}_i$ by calculating its hash function $h(\tilde{\mathbf{y}}_i)$, where $h$ is a homomorphic hash function. (The details of the homomorphic hash function which we use will be provided in Section \ref{sec:System Model}.) The master also calculates $h(B_i \mathbf{x})$ using its local information, \ie using $h(\mathbf{x})$ and $B_i$. 
If the master finds that $h(\tilde{\mathbf{y}}_i) \neq  h(B_i \mathbf{x})$, it concludes that the computed value is corrupted. Otherwise, $\tilde{\mathbf{y}}_i$ is declared as verified. 

\hfill $\Box$ 
\end{example}

The above example shows how homomorphic hash functions can be used for coded computation. However, existing hash-based solutions \cite{krohn2004fly, gkantsidis2006cooperative} introduce high computational overhead, which is not suitable for edge applications, where computation power and energy are typically limited. In this paper, we use homomorphic hash functions and coded computation gracefully and efficiently. In particular, we develop
and analyze light-weight and heavy-weight integrity check tools for coded computation using homomorphic hash functions. We design \SC by exploiting both light- and heavy-weight tools.
The following are the key contributions of this work:

\begin{itemize}
    \item We use a homomorphic hash function as in \cite{gkantsidis2006cooperative} and show that the hash of a linear combination of computed values can be constructed by the hashes of the original tasks. 
    \item We develop light- and heavy-weight integrity check tools for coded computation, and analyze these tools in terms of computation complexity and attack detection probability. We also analyze the trade-off between using  light- and heavy-weight tools for different  number of tasks.  
    \item We design \SC by exploiting light- and heavy-weight tools. If an attack is detected, \SC can pinpoint which tasks are corrupted. 
    \item We analyze the task completion delay of \SC by providing an upper bound as well as a lower bound on the gap between the task completion delay of \SC and a baseline.
    \item We evaluate \SC for different number and strength of malicious (Byzantine) workers. The simulation results show that our algorithm significantly improves task completion delay as compared to the baselines.  
\end{itemize}

The structure of the rest of this paper is as follows.
Section~\ref{sec:System Model} presents our system model.
Section~\ref{sec:LW-HW} presents light- and heavy-weight integrity check tools.
Section \ref{sec:SC} presents our secure coded cooperative computation (\SC) algorithm. 
Section \ref{sec:TrivialCompare} provides the theoretical analysis of the task completion delay of \SC.
Section \ref{sec:simulation} provides simulation results of \SC. 
Section \ref{sec:related} presents related work. 
Section \ref{sec:conc} concludes the paper.

\section{\label{sec:System Model} System Model}

{\em Setup.}  We consider a master/worker setup at the edge of the network, where the master device offloads its computationally intensive tasks to workers $w_n$, $n \in \Nset$ (where $|\Nset| = N$) via device-to-device (D2D) links such as Wi-Fi Direct and/or Bluetooth. The master device divides a task into smaller sub-tasks, and offloads them to parallel processing workers. 
{\em Task Model.} Our focus is on computation of linear functions; \ie the master device would like to compute the multiplication of matrix $A$ with vector $\mathbf{x}$; $y=A\mathbf{x}$, where $A = (a_{i,j})\in \mathbb{F}_{\psi}^{R \times \Co}$, $\mathbf{x} = (x_i) \in \mathbb{F}_{\psi}^{\Co \times 1}$, and $\mathbb{F}_{\psi}$ is a finite field. The motivation of focusing on linear functions stems from matrix multiplication applications where computing linear functions is a building block of several iterative algorithms such as gradient descent.  

{\em Coding.} We divide matrix $A$ into $R$ rows denoted by $A_i$, $i = 1, \ldots, R$. The master device applies Fountain coding \cite{LT, Raptor, Fountain}  across rows to create coded information packets $\mathbf{q}_j \triangleq \sum_{i=1}^{R} \gamma_{i,j}A_i$, $j=1, 2, \ldots, R+\epsilon$, where $\epsilon$ is the overhead required by Fountain coding\footnote{The overhead required by Fountain coding is typically as low as 5\% \cite{Fountain}.}, and $\gamma_{i,j} \in \{0,1\}$ are coding coefficients of Fountain coding and the information packet $\mathbf{q}_j$ is a row vector with size $\Co$. Rateless coding enabled by Fountain codes is compatible with our goal to deal with heterogeneity and time-varying nature of resources. In other words, coded packets are generated on the fly and transmitted to workers depending on the amount of their resources (as described in Section \ref{sec:C3P}) and Fountain codes are flexible to achieve this goal. 

{\em Worker \& Attack Model.}  The workers incur random delays while executing the task assigned to them by the master device. The workers have different computation and communication specifications resulting in a heterogeneous environment which includes workers that are significantly slower than others, known as stragglers. Moreover, the workers cannot be trusted by the master. In particular, we consider Byzantine attacks, where one or more workers can corrupt the tasks that are assigned to them.

{\em Homomorphic Hash Function.} We consider the following hash function that maps a large number $a$ to an output with much smaller size
\begin{align}\label{eq:hash}
 h(a) \triangleq \mod(g^{\mod(a,q)}, r),
\end{align}
where $q$ is a prime number selected randomly from the field $\mathbb{F}_{\phi}$, $r$ is a prime number that satisfies $q|(r-1)$ (\ie $r-1$ is divisible by $q$) and $g$ is a number in $\mathbb{F}_r$ which is calculated as $g=b^{(r-1)/q}$ for a random selection of $b \in \mathbb{F}_r, b\neq 1$ \cite{krohn2004fly,gkantsidis2006cooperative}. The defined hash function is a collision-resistant hash function with the property that when $\phi$ increases, $a$ is compressed less; \ie $h(a)$ becomes a better approximation of $a$ for larger $\phi$. However, the computational cost of calculating $h(a)$ increases for larger $\phi$. Thus, there is a trade-off between computational complexity and better approximation of $a$ in calculating $h(a)$. Our goal is to exploit this trade-off in the context of coded computation as described in the next sections. Another property of the defined hash function is homomorphism, \ie $h(\sum_i c_ia_i)=\prod_i h(a_i)^{c_i}$, which we will exploit in matrix-vector multiplication (in Section \ref{sec:LW-HW}). 

{\em Delay Model.}  Each packet transmitted from the master to a worker $w_n,\ n=1,2,...,N$ experiences the following delays: (i) transmission delay for sending the packet from the master to the worker, (ii) computation delay for computing the multiplication of the packet by the vector $\mathbf{x}$, and (iii) transmission delay for sending the computed packet from the worker $w_n$ back to the master. We denote by $\beta_{n,i}$ the computation time of the $i^\text{th}$ packet at worker $n$. 

\section{\label{sec:LW-HW}Light- and Heavy-Weight Integrity Check Tools for Coded Computation}
In this section we present how homomorphic hash functions considered in \cite{krohn2004fly, gkantsidis2006cooperative} and defined in (\ref{eq:hash}) are used gracefully with coded computation. We first show that (\ref{eq:hash}) can be applied to coded computation. Then, we develop light- and heavy-weight integrity check tools. The tools we develop in this section will be building blocks of our secure coded cooperative computation mechanism (\SC). 
\subsection{Homomorphic Hash Function for Coded Computation} 

Let us consider that $Z_n$ coded information packets are offloaded to worker $w_n$. The $i^\text{th}$ packet offloaded to $w_n$ is $\mathbf{p}_{n,i} \in \{\mathbf{q}_1, \ldots, \mathbf{q}_{R+\epsilon} \}$, which can be represented as $\mathbf{p}_{n,i} = \left(p_{n,i,1},\dots, p_{n,i,\Co} \right)$, where $p_{n,i,j}$ is the $j^\text{th}$ element of vector $\mathbf{p}_{n,i}$. Worker $w_n$ calculates $y_{n,i} = \mathbf{p}_{n,i} \mathbf{x}$ and sends it back to the master device. 

Assume that the master receives $\tilde{y}_{n,i}$ from $w_n$, where $\tilde{y}_{n,i} = y_{n,i}$ if packet is not corrupted. The master device checks the integrity of packets calculated at $w_n$ according to the following rule. First, it calculates 
\begin{align} \label{eq:alpha_n}
\alpha_n = h(\sum_{i=1}^{Z_n} c_i \tilde{y}_{n,i}),
\end{align} using the hash function defined in (\ref{eq:hash}), where $c_i$'s are coefficients (We will discuss how $c_i$ is selected later in this section.). Next, it calculates 
\begin{align} \label{eq:beta_n}
\beta_n = \mod\Big(\prod_{j=1}^{\Co} h(x_j)^{\mod\big((\sum_{i=1}^{Z_n} {c_{i}}p_{n,i,j}), q \big)}, r\Big),
\end{align} where $x_j$ is the $j^\text{th}$ element of vector $\mathbf{x}$, and $q$ and $r$ are the parameters of the hash function defined in (\ref{eq:hash}). $\beta_n$ in (\ref{eq:beta_n}) is calculated by the master device using its local data $\mathbf{p}_{n,i}$ and $\mathbf{x}$. $\beta_n$ is used to check $\alpha_n$ as described in the next theorem. 

\begin{theorem} \label{th:th1}
If $w_n$ does not corrupt packets, \ie $\tilde{y}_{n,i} = y_{n,i}$, $\forall i$, and $c_i$ is a nonzero integer, then $\alpha_n = \beta_n$ holds. 
\end{theorem} 
{\em Proof:}
The proof is provided in Appendix A. 
\hfill $\blacksquare$

We note that Theorem \ref{th:th1} is necessary, but not sufficient condition to determine if $w_n$ is malicious or not. The sufficiency condition depends on how $c_i$ is selected as explained next. 
\subsection{Light-Weight Integrity Check (LW Function)} 
The light-weight integrity check (LW function) uses Theorem \ref{th:th1} to determine if workers corrupt packets or not. In particular, LW function calculates $\alpha_n$ in (\ref{eq:alpha_n}) and $\beta_n$ in (\ref{eq:beta_n}) by selecting $c_i$ randomly and uniformly from $\{-1,1\}$. LW function concludes that packets processed by $w_n$ are not corrupted if $\alpha_n = \beta_n$. However, as we discussed earlier, this condition is not always a sufficient condition, so LW function detects  attacks with some probability, which is provided next. 

\subsubsection{Probability of Attack Detection} 
We first consider a pairwise Byzantine attack, where malicious worker $w_n$ corrupts two packets out of $Z_n$ packets by adding and subtracting terms. For example, $\tilde{y}_{n,i} = y_{n,i} + \delta_i$ and $\tilde{y}_{n,j} = y_{n,j} - \delta_j$, for any arbitrary $i$,$j\leq Z_n$ satisfying $i \neq j$. In this attack pattern, if $|\delta_i| \neq |\delta_j|$, and considering that the coefficients are selected from $\{-1,1\}$ in LW function, the attack is detected with 100\% probability. On the other hand, if the attack is symmetric, \ie $|\delta_i| = |\delta_j|$, the probability of detecting the attack is 50\%. As symmetrical attacks are the most difficult ones to detect, we focus on this scenario in the next lemma. 

\begin{lemma} \label{lemma:p_d_1}
Consider an attack where the malicious worker $w_n$ selects an even number $\tilde{Z}_n$ randomly out of $Z_n$ packets and corrupt them by adding $\delta$ to half of them, and subtracting $\delta$ from the other half. The probability of attack detection by LW function is 
\begin{align}\label{eq:p_d_1}
  P_{\text{detect}}^{\text{LW}} & = 
  %Pr_{F_{LW}}^d(Ic_n) & = 
  1-\Big(\frac{\tilde{Z}_n!}{2^{\tilde{Z}_n} \big(({\tilde{Z}_n}/{2})!\big)^2}\Big).
\end{align}
\end{lemma}
{\em Proof:}
The proof is provided in Appendix B.
\hfill $\blacksquare$  

As seen from Lemma \ref{lemma:p_d_1}, the probability of attack detection increases with increasing number of corrupted packets. This result intuitively holds for any attack pattern as the coefficients ($c_i$) are selected randomly for each packet and estimating these values by an attacker becomes difficult for larger set of corrupted packets. 
Another attack pattern and its detection probability are provided in the following. 

Consider an attack pattern where the malicious worker $w_n$ corrupts three packets out of $Z_n$ packets by adding $\delta$ to two of randomly selected computed packets and subtracting $2\delta$ from another randomly selected computed packet. This attack pattern can be detected unless the coefficients for the three corrupted packets are all $1$'s or all $-1$'s. Therefore, the probability of attack detection for this attack pattern is $(1-2/2^3)\times 100=75\%$. For a general attack pattern, the following lemma, provides a lower bound on the probability of attack detection.

\begin{proposition} \label{lemma:p_d_1_bound}
The probability of attack detection when LW function is used and for any attack pattern is lower bounded by $P_{\text{detect}}^{\text{LW}}  \geq 0.5$. 
\end{proposition}
{\em Proof:}
The proof is provided in Appendix C.
\hfill $\blacksquare$

\subsubsection{Computational Complexity} 

\begin{theorem} \label{th:complexity_LW}
The computational complexity of LW function for checking $Z_n$ packets calculated by $w_n$ is $O(\Co M(r) \log_2 q)$, where $\Co$ is the size of each information packet, $M(r)$ is the complexity of multiplication in $\mathbb{F}_r$, and $r$ and $q$ are the parameters of the hash function defined in (\ref{eq:hash}).  
\end{theorem}

{\em Proof:} The complexity of LW function consists of two parts; calculation of $\alpha_n$ in (\ref{eq:alpha_n}) and $\beta_n$ in (\ref{eq:beta_n}). We first analyze computational complexity of calculating $\alpha_n$. The sum $\sum_{i=1}^{Z_n} c_i \tilde{y}_{n,i}$ only has addition and subtraction as $c_i \in \{-1,1\}$, and can be ignored. The complexity of the modular exponentiation while calculating the hash function is $O(M(r)\log_2 q)$ by using the method of exponentiation by squaring. 

Similarly, we can calculate the computational complexity of calculating $\beta_n$. The complexity for computing $ \sum_{i=1}^{Z_n} c_{i}p_{n,i,j}$ corresponds to the complexity of addition and subtraction, which is negligible. The complexity of computing $\mod(\prod_{j=1}^{\Co} h(x_j)^{\mod(\sum_{i=1}^{Z_n} {c_{i}}p_{n,i,j},q)}, r)$ has two components: (i) Calculating the modular exponentiations $h(x_j)^{\mod(\sum_{i=1}^{Z_n} {c_{i}}p_{n,i,j},q)}, \forall j=1,2,...,\Co$: The complexity for this calculation is $O(M(r)\log_2 q)$ for one modular exponentiation and $O(\Co M(r)\log_2 q)$ for all $\Co$ modular exponentiations. (ii) Multiplying all the calculated modular exponentiations, \ie $\prod_{j=1}^{\Co} h(x_j)^{\mod(\sum_{i=1}^{Z_n} {c_{i}}p_{n,i,j},q)}$ in $\mathbb{F}_r$: The complexity for this calculation is $O((\Co-1) M(r))$. Thus, the total complexity of LW function becomes $O(\Co M(r) \log_2 q)$. This concludes the proof. \hfill $\blacksquare$

Noting that the computational complexity of calculating the original matrix multiplication is $O(R\Co M(\psi))$, where $M(\psi)$ is the complexity of multiplication in $\mathbb{F}_\psi$. As seen, the complexity of the LW function is significantly low, compared to the original task. This means LW function provides security check with low complexity. However, the probability of attack detection using LW function could be as low as 50\%, which may not be acceptable in some applications. Thus, we provide a heavy-weight integrity check tool (HW function) in the next section. Our ultimate goal is to use LW and HW functions together for higher attack detection probability while still having low computational complexity.

\subsection{Heavy-Weight Integrity Check (HW Function)} 
The heavy-weight integrity check (HW function) uses Theorem \ref{th:th1} similar to the LW function, but chooses the coefficients $c_i$ from a larger field $\mathbb{F}_{q}$ rather than $\{-1,1\}$. This selection, \ie choosing coefficients from a larger field, comes with larger attack detection probability and computational complexity as described next.  

\subsubsection{Probability of Attack Detection} 
\begin{lemma} \label{lemma:p_d_3}
The probability that HW function detects a Byzantine attack with any attack pattern is expressed as 
\begin{align} \label{eq:PdetectHW}
P_{\text{detect}}^{\text{HW}} = 1 - \frac{1}{q}    
\end{align} where $q$ is the parameter of the hash function in (\ref{eq:hash}). 
\end{lemma}
{\em Proof:}
The proof is provided in Appendix D.
\hfill $\blacksquare$

As seen from Lemma~\ref{lemma:p_d_3}, the attack detection probability increases with increasing $q$. Next, we present the computational complexity of HW function. 

\subsubsection{Computational Complexity} 
\begin{theorem} \label{th:complexity_HW}
The computational complexity of HW function for checking $Z_n$ packets calculated by $w_n$ is $O(\Co Z_n M(\phi))$. 
\end{theorem}
{\em Proof:}
The proof follows the same logic of the proof of Theorem \ref{th:complexity_LW}, \ie the complexity of HW function depends on calculating $\alpha_n$ in (\ref{eq:alpha_n}) and $\beta_n$ in (\ref{eq:beta_n}). The difference as compared to the proof of Theorem \ref{th:complexity_LW} is that $c_i$'s are selected from a larger field, so reducing multiplication to addition in $\sum_{i=1}^{Z_n} c_i \tilde{y}_{n,i}$ of (\ref{eq:alpha_n}) and $\sum_{i=1}^{Z_n} {c_{i}}p_{n,i,j}$ of (\ref{eq:beta_n}) cannot be made. In particular, the complexity of calculating these terms is $O(Z_n M(\phi))$. Following similar steps as in the proof of Theorem \ref{th:complexity_LW}, we can conclude that the computational complexity of HW function becomes $O(\Co M(r)\log_2 q) + O(\Co Z_n M(\phi))$. Since the second term dominates the computational complexity for large $R$ (hence $Z_n$), we calculate the computational complexity as $O(\Co Z_n M(\phi))$. This concludes the proof. \hfill $\blacksquare$

We can approximate $Z_n$ to $(R+\epsilon)/N$ on average assuming that coded information packets are distributed homogeneously across workers, where $R$ is the number of information packets, $\epsilon$ is the Fountain coding overhead, and $N$ is the number of workers. Thus, the computational complexity of HW function across all workers becomes $N O(\frac{\Co (R+\epsilon)}{N} M(\phi))$. As we discussed earlier, the computational complexity of the original matrix multiplication is $O(R\Co M(\psi))$. We also note that $M(\psi) >> M(\phi)$. This means that even though HW function is computationally-complex as compared to LW function, it is still computationally-efficient with respect to the original matrix multiplication (considering that $\epsilon$ is small and approaches to 0 with increasing number of packets).

\subsection{Light- versus Heavy-Weight Integrity Check} 
In this section, we investigate employing LW function multiple rounds/times to achieve higher attack detection probability with low computational complexity.  LW function is used to check $Z_n$ packets computed by $w_n$ by selecting $c_i$ uniformly randomly from $\{-1,1\}$. Let us call this the first round. In the second round, we can use LW function again, but selected values of $c_i$ will be different from the first round. Thus, if an attack is not detected in the first round, it may still be detected in the next round. Thus, using LW function over multiple rounds will increase the attack detection probability. The next theorem characterizes the performance of LW function when used in multiple rounds as compared to HW function. 

\begin{theorem} \label{th:compareLWHW}
The attack detection probability of multiple-round LW function is equal to the attack detection probability of HW function in (\ref{eq:PdetectHW}) when LW function is used for $\log_2(q)$ rounds. Furthermore, the computational complexity of $\log_2(q)$-round LW function is lower than HW function if the following condition is satisfied. 
\begin{align}\label{eq:function_a_vs_c}
  Z_n \geq \frac{M(r)}{M(\psi)}(\log_2 q)^2,
\end{align}
\end{theorem}
{\em Proof:}
The proof is provided in Appendix E.
\hfill $\blacksquare$ 

\section{\label{sec:SC} \SC: Secure Coded Cooperative Computation}
In this section, we present our secure coded cooperative computation (\SC) mechanism. \SC consists of packet offloading, attack detection, and attack recovery modules. 
\subsection{\label{sec:C3P} Dynamic Packet Offloading} 

The dynamic packet offloading module of \SC is based on \cite{KS18}. In particular,  the master offloads coded packets gradually to workers and receives two ACKs for each transmitted packet; one confirming the receipt of the packet by the worker, and the second one (piggybacked to the computed packet) showing that the packet is computed by the worker. Then, based on the frequency of the received ACKs, the master decides to transmit more/less coded packets to that worker. In particular, each packet $\mathbf{p}_{n,i}$ is transmitted to each worker $w_n$ before or right after the computed packet $\mathbf{p}_{n,i-1}\mathbf{x}$ is received at the master. For this purpose, the average per packet computing time $E[\beta_{n,i}]$ is calculated for each worker $w_n$ dynamically based on the previously received ACKs. Each packet $\mathbf{p}_{n,i}$ is transmitted after waiting $E[\beta_{n,i}]$ from the time $\mathbf{p}_{n,i-1}$ is sent or right after packet $\mathbf{p}_{n,i-1}\mathbf{x}$ is received at the master, thus reducing the idle time at the workers. This policy is shown to approach the optimal task completion delay and maximizes the workers' efficiency and is shown to improve task completion delay significantly compared with the literature \cite{KS18}.

\subsection{Attack Detection} 
Assume that while the dynamic packet offloading process continues, the set of received packets from worker $w_n$ at the master device during time interval $T$ is $\Zset_n$ ($|\Zset_n|=Z_n$). The attack detection module of \SC is applied on $\Zset_n$ periodically and consists of two phases. 

The first phase applies LW function on the packets in $\Zset_n$ for any worker $w_n, n \in \Nset$. Let us assume that attack is detected in the packets coming from worker $w_{n^*}$. Then, all the packets in $\Zset_{n^*}$ are discarded and the malicious worker $w_{n^*}$ is removed from the set of workers, \ie $\Nset = \Nset - n^*$. As we discussed earlier, the attack detection probability of LW function increases with increasing corrupted packets. Thus, if an attack is detected in this phase, we can consider that most of the packets are corrupted, so we can discard all the received packets. 

The goal of the second phase is to detect any attacks, which are not detected in the first phase. Both HW and multiple-round LW functions are used in this phase. In particular, if the inequality in Theorem \ref{th:compareLWHW} is satisfied, LW function is used for $\log_2(q)$ times. Otherwise, HW function is used. If an attack is not detected, all the packets in $\Zset_n$ are labeled as {\em verified} packets. Otherwise, \ie if an attack is detected, the attack recovery module, which is described later in this section, starts. 

\subsection{Attack Recovery}
If an attack is detected in the second phase of the attack detection module of \SC, we consider that a small number of packets are corrupted. Otherwise, the first phase of the attack detection module could have detected the attack and discarded all the packets. Thus, the goal of the attack recovery module is to detect a small number of corrupted packets and recover the non-corrupted packets, \ie avoid discarding all the packets. 

Let us assume that an attack is detected among the packets received from $w_{n^*}$, \ie in $\Zset_{n^*}$. In order to pinpoint the packets that are corrupted, we use a binary search algorithm. In particular, $\Zset_{n^*}$ is divided into two disjoint sets; $\Zset_{n^*}^1$ and $\Zset_{n^*}^2$. The second phase of the attack detection module is run over these two sets. If an attack is not detected on any of these sets, all the packets in that set are verified. Otherwise, the binary search (this set splitting) continues over the sets where an attack is detected. When the size of a splitted set is one, \ie it has one packet in it, and an attack is detected, the packet in that set is declared a corrupted packet and discarded. As seen, the attack recovery module can still verify some of the packets coming from a malicious worker. This is important to efficiently utilize available resources while still providing security guarantees. 

\subsection{\SC in a nutshell}
\SC algorithm is provided in Algorithm \ref{alg:SC3P}. As detailed in this algorithm, in \SC, the attack detection module, which if required will be followed by the attack recovery module, is applied until the number of verified packets from all workers reaches $R + \epsilon$ (Note that $R + \epsilon$ is the number of packets required for a successful decoding of Fountain codes). In particular, first, the attack detection module is applied on each $\Zset_n$ set of packets received from worker $w_n, n \in \Nset$ during the time period $T$, where $T$ is defined as the time period that $R+\epsilon$ packets are received collectively from all workers, \ie $\sum_{n=1}^N Z_n = R+\epsilon$. If all $R+\epsilon$ packets in the set $\cup_{n=1}^N \Zset_n$ are labeled as {\em verified}, then all workers have been honest and sent back correct results to the master device. Otherwise, \ie if an attack is detected, then the number of correct packets delivered by honest workers and labeled as {\em verified}, is less than $R+\epsilon$. In this case, the master device waits until it receives additional $R+\epsilon-V$ packets collectively from all workers, where $V$ is the number of packets labeled as {\em verified}. Then, for each worker $w_n, n\in\Nset$, the attack detection module is applied on the set of newly received packets. This process is repeated until $R+\epsilon$ packets are labeled as {\em verified}. Finally, Fountain decoding is applied on $R+\epsilon$ packets labeled as {\em verified} and the result of the multiplication task is obtained by the master device.

\begin{algorithm}[t!]
\caption{\SC}
\label{alg:SC3P}
\begin{algorithmic}[1]
\STATE $V=0$
\WHILE{$V < R+\epsilon$}
    \STATE Determine the time period $T$ as the time interval during which $R+\epsilon -V$ computed packets are received from all workers collectively. 
    \FOR{$n=1:N$}
        \STATE Create the set $\Zset_n$ consisting of packets received from worker $w_n$ during the time period $T$.
        \STATE $V_{add}=0$.
        \STATE Apply the attack detection module on $\Zset_n$ and set $V_{add}$ as the number of packets labeled as {\em verified}.
        \STATE Update $V$ as $V+V_{add}$.
        \IF {$V \geq R+\epsilon$}
            \STATE Stop the process and use $R+\epsilon$ packets labeled as {\em verified} for Fountain decoding. 
        \ENDIF
    \ENDFOR
\ENDWHILE
\end{algorithmic}
\end{algorithm}

\section{\label{sec:TrivialCompare}Performance Analysis of \SC}
In this section, we first characterize the task completion delay of \SC and then we provide a lower bound on the gap between the task completion delay of \SC and the task completion delay of a baseline. The task completion delay is the time spent to receive $R+\epsilon$ computed and verified packets at the master device collectively from all workers.

\begin{theorem} \label{th:T_SC3P}
The average task completion delay of \SC for a set of workers $\{w_n, n\in\Nset\}$, out of which $\{w_n, n\in\Nset_m\}$ is the set of malicious workers, is upper bounded by:
\begin{align}\label{eq:T_SC3P}
  &E[T_{\text{SC}^3}] \leq\\
  &\frac{R+\epsilon}{\sum_{n\in\Nset}1/E[\beta_{n,i}]}+\frac{\sum_{n \in \Nset_m}z_n \Big(P+ \rho_c\big(1-P\big)\Big)}{\sum_{n \notin \Nset_m}1/E[\beta_{n,i}]},
\end{align}
where $z_n$ is equal to $$z_n=\frac{R+\epsilon}{E[\beta_{n,i}]\sum_{n \in \Nset} 1/E[\beta_{n,i}]},$$ $\rho_c$ is the probability that a packet is corrupted by a malicious worker, and $P$ is given by $$P=1-\Big(\frac{(z_n\rho_c)!}{2^{(z_n\rho_c)} \big(({z_n\rho_c}/{2})!\big)^2}\Big).$$
\end{theorem}
{\em Proof:}
The proof is provided in Appendix F.\hfill $\blacksquare$

In the following we characterize the task completion delay of \SC as compared with a baseline, where the master detects the malicious workers and takes advantage of only honest workers to accomplish its task. One method to detect the malicious workers is using HW function with a high value for the parameter $q$, so that the probability of attack detection given in (\ref{eq:PdetectHW}) is close to $1$. We call this baseline as HW-only and denote its task completion delay by $T_{\text{HW-only}}$. Note that in HW-only, if a worker is detected as malicious, all the packets coming from that worker are discarded, while \SC uses both LW and HW functions gracefully to discard only corrupted packets coming from malicious workers.

\begin{lemma} \label{lemma:T_trivial-T_SC3P}
The gap between the task completion delay of HW-only and the task completion delay of \SC is lower bounded by:
\begin{align}\label{eq:T_trivial-T_SC3P}
  &T_{\text{HW-only}}-E[T_{\text{SC}^3}] \ge\\ 
  &\frac{(R+\epsilon) (1-\rho_c) \sum_{n \in \Nset_m}\frac{1-P}{E[\beta_{n,i}]}}{\big(\sum_{n \in \Nset}1/E[\beta_{n,i}]\big) \big(\sum_{n \notin \Nset_m}1/E[\beta_{n,i}]\big)}.
\end{align}
\end{lemma}
{\em Proof:}
The proof is provided in Appendix G. \hfill $\blacksquare$

From Lemma \ref{lemma:T_trivial-T_SC3P}, we can conclude that the faster the honest helpers are, the closer are the performances of HW-only and our \SC. This is expected as the performance of \SC is dominated by the fastest workers and the performance of HW-only is dominated by the speed of honest workers and thus \SC performs close to HW-only when the fastest workers are honest. In addition, smaller $\rho_c$ (which results in smaller $P$) results in larger gap between HW-only and \SC. This is expected as smaller $\rho_c$ results in less number of corrupted packets delivered by malicious workers and thus using \SC that takes advantage of non-corrupted packets delivered by malicious workers results in more performance improvement compared with HW-only that throws away non-corrupted packets delivered by malicious workers. Finally, lower bound on the gap $T_{\text{HW-only}}-E[T_{\text{SC}^3}]$ is linearly proportional to $R+\epsilon$. This implies that more improvement is obtained by using \SC compared with HW-only for larger input matrix $A$.

\section{\label{sec:simulation} Performance Evaluation}
In this section, we evaluate the performance of our algorithm; Secure Coded Cooperative Computation (\SC) via simulations. We consider master/worker setup, where some of the workers are malicious. Each computed packet $y_{n,i}$ is corrupted by the malicious worker $w_n$ with probability $\rho_c$. The computing resources are heterogeneous and vary across workers, where per packet computing delay $\beta_{n,i}$ is an i.i.d. random variable following a shifted exponential distribution. We compare \SC with the baselines (i) HW-only, which uses HW function to detect corrupted packets, while \SC uses both LW and HW functions gracefully. In HW-only, if a worker is detected as malicious, all the packets coming from that worker are discarded, (ii) Lower Bound, which is obtained by using C3P proposed in \cite{KS18}, the best known dynamic but unsecured coded cooperative computation. Note that C3P is not practical in the presence of an attacker, however it can provide a lower bound on \SC and the gap between Lower Bound and \SC shows the cost that we should pay to make our system secure against Byzantine attack, and (iii) Upper Bound provided in (\ref{eq:T_SC3P}).

\begin{figure}[t!]
\centering
{\includegraphics[height=50mm]{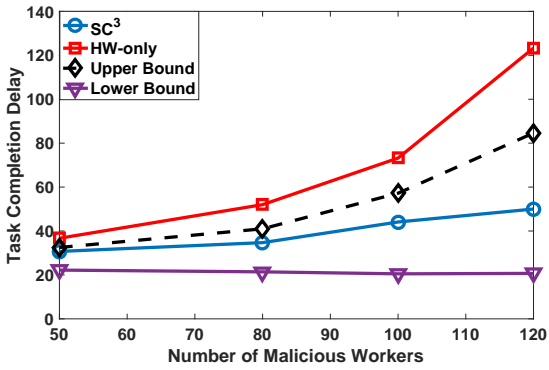}} 
\caption{Task completion delay of \SC as compared to (i) HW-only, (ii) Lower Bound, and (iii) Upper Bound, with increasing number of malicious workers.}
\label{fig:CT_numberOfMalWorkers}
\end{figure}

{\em Task Completion Delay vs. Number of Malicious Workers.} Fig. \ref{fig:CT_numberOfMalWorkers} compares the task completion delay of \SC with the baselines for increasing number of malicious workers. 

In this setup, the total number of workers is $N=150$, the number of rows in matrix $A$ is $R = 1K$, the number of columns is $\Co = 1K$, the overhead of Fountain codes  is $5\%$, the probability of packet corruption is $\rho_c = 0.3$, and per-packet computing delay is a shifted exponential random variable with the mean selected uniformly between $1$ and $6$ for each worker. 

The task completion delay of \SC and HW-only increases with increasing number of malicious workers. When the number of malicious workers increases, there will be more corrupted packets in the system. These corrupted packets are detected and discarded by \SC and HW-only. As more packets are discarded when the number of malicious workers is higher, the task completion delay increases. The increase in the task completion delay of \SC is less than HW-only thanks to (i) using both LW and HW functions to reduce completion time and thus computational complexity, and (ii) attack recovery module of \SC. \SC performs better than its Upper Bound as the Upper Bound is based on the theoretical analysis in the worst case scenario. Finally, the completion time of Lower Bound does not change by increasing the number of malicious workers as it uses C3P in \cite{KS18}, which is not designed for an environment with malicious workers and uses all received packets including the corrupted packets to obtain the computation task result. By increasing the number of malicious workers, the gap between the performance of \SC and the Lower Bound increases, as the cost for providing a secure system increases when the adversary attacks more workers.

\begin{figure}[t!]
\centering
{\includegraphics[height=50mm]{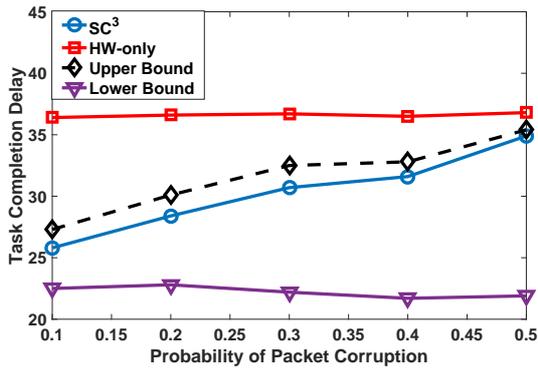}} 
\caption{Task completion delay of \SC as compared to (i) HW-only, (ii) Lower Bound, and (iii) Upper Bound, with increasing probability of packet corruption.}
\label{fig:CT_probPacketCorruption}
\end{figure}

{\em Task Completion Delay versus Packet Corruption Probability.} Fig. \ref{fig:CT_probPacketCorruption} compares the task completion delay of \SC with (i) HW-only, (ii) Lower Bound, and (iii) Upper Bound for different values of $\rho_c$, the probability that a delivered packet by a malicious worker is corrupted. The number of workers, the number of rows in $A$, Fountain coding overhead, and per-packet computing delay are the same as the previous setup above. The number of malicious workers is $N_m=50$. 

The task completion delay of HW-only does not change with increasing packet corruption probability. The reason is that HW-only does not have attack recovery feature and discards all the packets coming from a malicious worker. On the other hand, task completion delay of \SC is significantly lower than HW-only especially when the packet corruption probability is low thanks to using both LW and HW functions and employing the attack recovery module. Again, the completion time of Lower Bound does not change by increasing $\rho_c$ and by increasing $\rho_c$, the gap between the performance of \SC and the Lower Bound increases, as the cost for providing a secure system increases when number of corrupted packets increases.

\begin{figure}[h!]
\centering
\subfigure[
Gap between HW-only and \SC vs. speed of computation at honest workers]{ \scalebox{.3}{\includegraphics{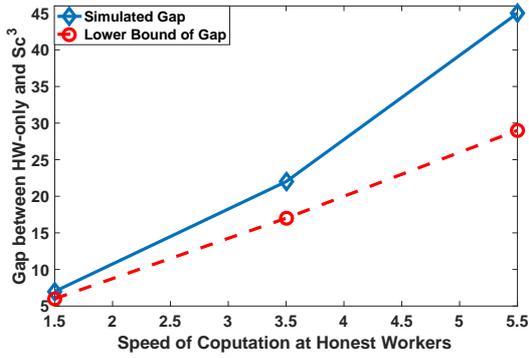}} }
\subfigure[
Gap between HW-only and \SC vs. the probability of packet corruption]{ \scalebox{.3}{\includegraphics{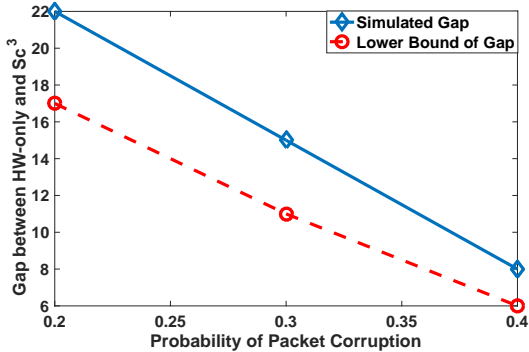}} }
\subfigure[
Gap between HW-only and \SC vs. number of rows of matrix $A$]{ \scalebox{.3}{\includegraphics{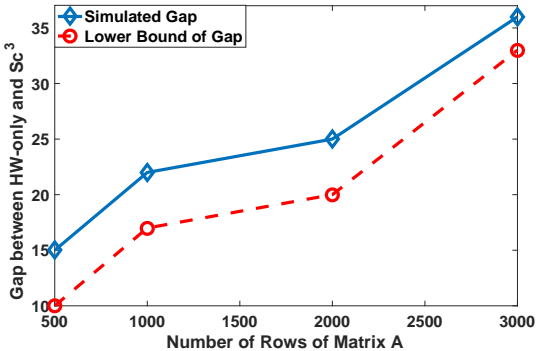}} }
\caption{
Gap between HW-only and \SC}
\label{fig:gap}
\end{figure}

{\em Task Completion Delay Gap between \SC and HW-only.} Fig. \ref{fig:gap} shows the gap between the HW-only and \SC and compares the simulated gap with the lower bound of the gap provided in (\ref{eq:T_trivial-T_SC3P}) for the total number of $N=80$ workers out of which $N_m=40$ are malicious. The number of rows in $A$ is $R=1K$ for Figs. \ref{fig:gap}(a) and (b), the number of columns is $\Co = 1K$, Fountain coding overhead is $5\%$, the probability of packet corruption is $\rho_c = 0.3$ for Figs. \ref{fig:gap}(a) and (c), and per-packet computing delay is a shifted exponential random variable with the mean selected uniformly between $3$ and $4$ for each worker for Figs. \ref{fig:gap}(b) and (c).

Fig. \ref{fig:gap}(a) shows the gap versus the speed of computation at honest helpers. The per-packet computing delay is a shifted exponential random variable with the mean selected uniformly between $3$ and $4$ for each malicious worker. For each honest worker, the mean is selected uniformly between $1$ and $2$ for the first simulated points, between $3$ and $4$ for the second simulated points, and between $5$ and $6$ for the third simulated points.
As seen, the faster the honest workers are, the closer are the performances of the HW-only and our \SC. This observation confirms our analysis in section \ref{sec:TrivialCompare}.

Fig. \ref{fig:gap}(b) shows the gap versus $\rho_c$, the probability of packet corruption by a malicious worker. As seen, larger $\rho_c$ (which results in more corrupted packets delivered by malicious workers) results in smaller gap between HW-only and \SC. This observation confirms our analysis in section \ref{sec:TrivialCompare}.

Fig. \ref{fig:gap}(c) shows the gap versus the number of rows of matrix $A$. As seen, the gap between HW-only and SC3P increases with an increase in the number of rows of matrix $A$. This observation confirms our analysis in section \ref{sec:TrivialCompare} stating that more improvement will be obtained by using \SC compared with HW-only for larger input matrix $A$.

\section{\label{sec:related} Related Work}

Coded computation, advocating mixing data in computationally intensive tasks by employing erasure codes and offloading these tasks to other devices for computation, has recently received a lot of attention,   \cite{KS18,BPR17,li2016unified,dutta2017coded,yang2017computing,halbawi2017improving,yu2017polynomial,DCG16,tandon2017gradient,li2016fundamental,lee2018speeding,karakus2017straggler,aktas2017effective}. For example, coded cooperative computation is shown to provide higher reliability, smaller delay, and reduced communication cost in MapReduce framework \cite{MapReduce}, where computationally intensive tasks are offloaded to distributed server clusters \cite{CodedMapReduce}. The effectiveness of coded computation in terms of task completion delay has been investigated in \cite{lee2018speeding,yu2017polynomial,KS18}. In \cite{reisizadeh2019coded}, the same problem is considered, but with the assumption that workers are heterogeneous in terms of their resources. In \cite{KS18}, a dynamic and adaptive algorithm with reduced task completion time is introduced for heterogeneous workers. As compared to this line of work, we consider secure coded computation by focusing on Byzantine attacks. 

There is existing work at the intersection of coded computation and security by specifically focusing on privacy \cite{BPR17,8382305,YRSA18, bitar2019prac}. As compared to this line of work, we focus on Byzantine attacks and use homomorphic hash functions. Homomorphic hash functions have been widely used for transmission of network coded data. Corrupted network coded packets are detected by applying homomorphic hash functions that we consider in this work \cite{krohn2004fly}. The hash function is applied to random linear combinations of network coded packets in \cite{gkantsidis2006cooperative}. \SC, although similar to these work, is more efficient in terms of computational efficiency, which was not the main concern of \cite{krohn2004fly, gkantsidis2006cooperative} as their focus was on transmitting network coded packets, not computation.   

\section{\label{sec:conc} Conclusion}

In this paper, we focused on secure edge computing against Byzantine attacks. We considered a master/worker scenario where honest and malicious workers with heterogeneous resources are connected to a master device. We designed a secure coded cooperative computation mechanism (\SC) that  provides both  security  and  computation  efficiency guarantees by gracefully combining homomorphic hash functions, and coded cooperative computation. Homomorphic hash functions are used against Byzantine attacks and  coded cooperative computation is used to improve computation efficiency when edge resources are heterogeneous and time-varying. Simulations results show that \SC improves task completion delay significantly.

\bibliographystyle{IEEEtran}
\bibliography{main}

\section*{Appendix A: Proof of Theorem~\ref{th:th1}}
\begin{align}\label{eq:proof_homo}
  \alpha_n&=h(\sum_{i=1}^{Z_n} c_i \tilde{y}_{n,i})\\
  & = \mod\big(g^{\mod(\sum_{i=1}^{Z_n} c_i \tilde{y}_{n,i}, q)},r\big) \\
  & = \mod\big(g^{\mod(\sum_{i=1}^{Z_n} c_i y_{n,i}, q)},r\big) \\
  & = \mod\big(g^{\mod(\sum_{i=1}^{Z_n} c_i \sum_{j=1}^{C} p_{n,i,j} x_j, q)}, r\big) \\
  & = \mod\big(g^{(\sum_{i=1}^{Z_n} \sum_{j=1}^{C} c_ip_{n,i,j} x_j)-q q'}, r\big) \\
  & = \mod\big(g^{\sum_{i=1}^{Z_n} \sum_{j=1}^{C} c_ip_{n,i,j} x_j}, r\big) \nonumber\\
  & \quad \quad \times \mod(g^{-q q'}, r), \label{eq:proof_homo_4}
\end{align}
where $q'$ is the quotient of dividing $(\sum_{i=1}^{Z_n} \sum_{j=1}^{C} c_ip_{n,i,j} x_j)$ by $q$. On the other hand, as mentioned before, $g$ can be written as $b^{(r-1)/q}$ and thus by using the condition of $q|(r-1)$ and using Fermat's little theorem, we have:
\begin{align}\label{eq:proof_homo_5}
  \mod(g^{q\times q'}, r) 
  & = \mod(b^{ q \times q' \times (r-1)/q},r) \\
  & = \mod(b^{(r-1) \times q'},r) \\
  & = 1. \label{eq:proof_homo_9}
\end{align}

From (\ref{eq:proof_homo_9}) and (\ref{eq:proof_homo_4}), we have:
\begin{align}
&\mod\big(g^{\mod(\sum_{i=1}^{Z_n} \sum_{j=1}^{C} c_ip_{n,i,j} x_j, q)}, r\big)\\
&=\mod\big(g^{\sum_{i=1}^{Z_n} \sum_{j=1}^{C} c_ip_{n,i,j} x_j}, r\big)
\end{align}
From the above equation, we can conclude that: 
\begin{align}
    \mod(g^{x_j},r)=\mod(g^{\mod(x_j, q)},r)\label{eq:1}
\end{align}
and
\begin{align}
    &\mod(g^{x_j\sum_{i=1}^{Z_n} c_ip_{n,i,j}},r)=\nonumber\\
    &\mod(g^{x_j\mod(\sum_{i=1}^{Z_n} c_ip_{n,i,j},q)},r)\label{eq:2}
\end{align}
On the other hand, from (\ref{eq:proof_homo_9}) and (\ref{eq:proof_homo_4}), we have:
\begin{align}\label{eq:proof_homo_10}
  \alpha_n
  & = \mod \big(g^{\sum_{i=1}^{Z_n} \sum_{j=1}^{C} c_ip_{n,i,j} x_j}, r\big)\\
  & = \mod(\prod_{j=1}^{C} g^{\sum_{i=1}^{Z_n} c_ip_{n,i,j} x_j},r) \\
  & = \mod(\prod_{j=1}^{C} g^{x_j\mod(\sum_{i=1}^{Z_n} c_ip_{n,i,j},q)},r)\label{eq:3}\\
  & = \mod(\prod_{j=1}^{C}\nonumber\\
  &\quad \quad (\mod(g^{\mod(x_j,q)},r))^{\mod(\sum_{i=1}^{Z_n} c_ip_{n,i,j},q)},r)\label{eq:4}\\
  & = \mod\big(\prod_{j=1}^{C} h(x_j)^{\mod(\sum_{i=1}^{Z_n} c_ip_{n,i,j}, q)}, r\big)\\
  & = \beta_n
\end{align}
where, (\ref{eq:3}) comes from (\ref{eq:2}) and (\ref{eq:4}) comes from (\ref{eq:1}). 
This concludes the proof.

\section*{Appendix B: Proof of Lemma~\ref{lemma:p_d_1}}
This kind of attack is not detected by the master device if the coefficients $c_i$ for corrupted packets are selected such that the added $\delta$'s are canceled out with the subtracted $\delta$'s. For example, let us assume there are $\tilde{Z}_n=4$ corrupted packets of $y_{n,1}+\delta, y_{n,2}+\delta, y_{n,3}-\delta,$ and $y_{n,4}-\delta$ among $Z_n=10$ packets received from worker $w_n$. For this attack not to be detected, there are six possibilities for values of coefficients $\{c_1, c_2, c_3, c_4\}$: $\{1,1,1,1\}$, $\{1,-1,1,-1\}$, $\{1,-1,-1,1\}$, $\{-1,1,1,-1\}$, $\{-1,1,-1,1\}$, and $\{-1,-1,-1,-1\}$. Note that the other six coefficients $\{c_5, \ldots, c_{10}\}$ can have any value as those packets are not corrupted. All possible cases for the first four coefficients are $2^4=16$ cases. Therefore, the probability of attack detection for this example is $1-6/16$. In general, for $\tilde{Z}_n$ corrupted packets, the number of cases the attack cannot be detected is equal to the combination of ${\tilde{Z}_n}\choose {\tilde{Z}_n/2}$. The reason is that half of the $\tilde{Z}_n$ coefficients for which the corrupted packets are added by $\delta$ can have any value but the other half for which the corrupted packets are subtracted by $\delta$ should be chosen such that when they are multiplied by their correspondent coefficients, the added $\delta$'s can be canceled out with the subtracted $\delta$'s. 
Note that any permutation of coefficients for which the corrupted packets are added by $\delta$ (or subtracted by $\delta$) do not have any effect on the attack detection. 
As the total number of cases for coefficient selections of the corrupted packets is equal to $2^{\tilde{Z}_n}$, the probability that the attack is not detected is equal to $\frac{\binom{\tilde{Z}_n}{\tilde{Z}_n/2}}{2^{\tilde{Z}_n}}=(\frac{\tilde{Z}_n!}{2^{\tilde{Z}_n}\times ((\frac{\tilde{Z}_n}{2})!)^2})$. This concludes the proof. 

\section*{Appendix C: Proof of Proposition~\ref{lemma:p_d_1_bound}}
If the malicious worker corrupts only one packet out of $Z_n$ packets, it is obvious that this attack can be detected by applying $F_{LW}$. Among the remaining attack patterns, \ie all attack patterns, where the malicious worker corrupts more than one packet, the most difficult one to detect is a symmetric pairwise Byzantine attack, where the malicious worker corrupts two packets out of $Z_n$ packets. The reason is that the attack is detected by applying function $F_{LW}$ unless the coefficients corresponding to the corrupted packets have a systematic structure, while the coefficients corresponding to the remaining packets can have any value. Therefore, for $F_{LW}$ to fail the attack detection, among all attacks with more than two corrupted packets, the function has the least freedom on selecting the coefficients when the number of corrupted packets is two. This results in the least probability of attack detection when the number of corrupted packets is two. This fact can also be confirmed in Lemma \ref{lemma:p_d_1}, as the detection probability presented in (\ref{eq:p_d_1}) is an increasing function of the number of corrupted packets. On the other hand, among all attack patterns that changes two of the packets, the symmetrical attacks, where one of the packets is corrupted by adding $\delta$ to the result and the other packet is corrupted by subtracting the same amount of $\delta$ from the result is the most difficult one to be detected (In fact, the probability of attack detection for the asymmetrical pairwise attack is $100\%$). Therefore, the most difficult attack pattern for function $F_{LW}$ is symmetrical pairwise Byzantine attack, for which the probability of attack detection is $50\%$. This concludes the proof. 

\section*{Appendix D: Proof of Lemma~\ref{lemma:p_d_3}}
In order for an attack not to be detected by applying HW function, the value of the corrupted packets should be changed by the attacker such that $\alpha_n$ in (\ref{eq:alpha_n}) is equal to $\beta_n$ in (\ref{eq:beta_n}). For this condition to be satisfied, $\mod(\sum_{i=1}^{Z_n} c_i \tilde{y}_{n,i},q)$ should be equal to $\mod(\sum_{i=1}^{Z_n} c_i y_{n,i},q)$. In other words, if the following condition is satisfied, then the attack will not be detected:
\begin{align}
    \mod(\sum_{i\in \tilde{\Zset}_n} c_i (y_{n,i}-\tilde{y}_{n,i}),q)=0, 
\end{align}
where $\tilde{\Zset}_n$ with size $|\tilde{\Zset}_n|=\tilde{Z}_n$ is the set of corrupted packets among all $Z_n$ received packets, \ie $(y_{n,i}-\tilde{y}_{n,i}) \neq 0, i\in \tilde{\Zset}_n$.
For this condition to be satisfied, one of the coefficients $\{c_j|j\in \tilde{\Zset}_n\}$ out of all $\tilde{Z}_n$ coefficients, should be selected depending on the values of the other coefficients, \ie $c_j$ should be selected such that the following condition is satisfied:
\begin{align}
    &\mod( c_j (y_{n,j}-\tilde{y}_{n,j}),q)=\nonumber\\
    &\mod(\sum_{{i\in \tilde{\Zset}_n},{i\neq j}} c_i (y_{n,i}-\tilde{y}_{n,i}),q)
\end{align}
Since $q$ is a prime number, from the modular arithmetic principles, $c_j$ has a unique solution. Considering that $c_j$ is selected randomly in $\mathbb{F}_{q}$ by the master device, the probability that the selected coefficient $c_j$ satisfies the above equation is $1/q$. Therefore, the probability that the attack is not detected by HW function is $1/q$. This concludes the proof.

\section*{Appendix E: Proof of Theorem~\ref{th:compareLWHW}}
According to Lemma \ref{lemma:p_d_1_bound}, the probability of attack detection when LW function is used for one round is at least $\frac{1}{2}$. When LW function is used for two rounds, \ie no attack is detected by selecting the coefficients $c_i, 1 \leq i \leq Z_n$ uniformly randomly from $\{-1,1\}$ and thus a different set of coefficients $c_i\in\{-1,1\}, 1 \leq i \leq Z_n$ are selected uniformly at random, the probability of attack detection is at least $1-\frac{1}{2} \times \frac{2^{Z_n-1}-1}{2^{Z_n}-1}$.
Similarly, the probability of attack detection when LW function is used for $K$ rounds is at least $1-\prod_{k=0}^{K} \frac{2^{Z_n-1}-k}{2^{Z_n}-k}$, which can be approximated as $\frac{1}{2^K}$ when $Z_n \gg \log_2 K$. Therefore, for $K=\log_2 q$, the probability of attack detection when LW function is used for $\log_2 q$ rounds is $1-1/q$, which is equal to the attack detection probability of HW function. 

According to Theorem \ref{th:complexity_LW}, the computational complexity for one round of LW function is $O(\Co M(r) \log_2 q)$ and thus the computational complexity of $\log_2 q$ rounds of LW function is $O(\Co M(r) (\log_2 q)^2)$. On the other hand, according to Theorem \ref{th:complexity_HW}, the computational complexity of HW function is $O(\Co Z_n M(\phi))$. Therefore, if $Z_n \geq \frac{M(r)}{M(\psi)}(\log_2(q))^2$, the computational complexity of $\log_2(q)$-round LW function is lower than HW function.

This concludes the proof. 

\section*{Appendix F: Proof of Theorem~\ref{th:T_SC3P}}
In order to characterize the completion time of \SC, we calculate the required time for receiving the required number of packets collectively from all workers at the master device during each time period $T$, defined in Algorithm \ref{alg:SC3P}.

According to Algorithm \ref{alg:SC3P}, the first time period $T$ is defined as the time interval during which $R+\epsilon$ packets are received collectively from all workers. Using the dynamic packet offloading module of \SC, this time period is equal to $\frac{R+K}{\sum_{n\in \Nset}1/E[\beta_{n,i}]}$, according to (17) in \cite{KS18}, where $E[\beta_{n,i}]$ is the average of per packet computing time at worker $w_n, n \in \Nset$. 

According to Algorithm \ref{alg:SC3P}, the second time period $T$ is defined as the time interval during which $R+\epsilon-V$ packets are received collectively from all workers, where $V$ is the number of packets labeled as {\em verified} after applying the attack detection module on the packets received during the first time interval. In the worst case scenario, $R+\epsilon-V$ additional packets that should be received at the master device and labeled as {\em verified}, are delivered only by honest workers. This worst case scenario results in the maximum time for receiving $R+\epsilon-V$ additional packets, which is equal to $\frac{R+\epsilon-V}{\sum_{n\notin \Nset_m}1/E[\beta_{n,i}]}$ \cite{KS18}. By taking into account this worst case scenario and the upper bound on the average value of $R+\epsilon-V$ (provided in Lemma \ref{lemma:Unverified_pack}), and adding the time during the first time period, the completion time is upper bounded by $\frac{R+K}{\sum_{n\in \Nset}1/E[\beta_{n,i}]}+\frac{\sum_{n\in\Nset_m} z_n (P+\rho_c(1-P))}{\sum_{n\notin \Nset_m}1/E[\beta_{n,i}]}$.

\begin{lemma} \label{lemma:Unverified_pack}
The average number of packets among all $R+\epsilon$ packets received during the first time period $T$, that are not labeled as {\em verified} by \SC, is upper bounded by:
\begin{align}\label{eq:Unverified_pack}
  R+\epsilon-V \leq \sum_{n\in\Nset_m} z_n (P+\rho_c(1-P)),
\end{align}
where $P$ is given by $P=1-\Big(\frac{(z_n\rho_c)!}{2^{(z_n\rho_c)} \big(({z_n\rho_c}/{2})!\big)^2}\Big)$.
\end{lemma}
{\em Proof:} The packets received during the first time period $T$, that are not labeled as {\em verified} by \SC, consist of two kinds:

(i) Packets received from malicious workers, where attack is detected by applying the LW function: The average number of these packets is equal to $\sum_{n \in \\Nset_m} z_n P$, where $P$ is equal to the probability of attack detection by LW function when applied on $z_n$ packets received from worker $w_n, n \in \Nset$. From (\ref{eq:p_d_1}), $P$ is given by $P=1-\Big(\frac{(z_n\rho_c)!}{2^{(z_n\rho_c)} \big(({z_n\rho_c}/{2})!\big)^2}\Big)$.

(ii) Corrupted packets received from malicious workers, where attack is not detected by applying the LW function but attack is detected in the attack recovery module. The average number of such packets is upper bounded by $\sum_{n \in \Nset_m} z_n \rho_c (1-P)$, where $z_n \rho_c$ is the average number of corrupted packets received from the malicious worker $w_n$ and $1-P$ is the probability that the attack is not detected by applying function LW. Note that for larger values of $q$, the probability of attack detection by applying HW function or multiple-round LW function is closer to 1 and the exact value gets closer to its upper bound.

This concludes the proof.
\hfill $\blacksquare$

\section*{Appendix G: Proof of Lemma~\ref{lemma:T_trivial-T_SC3P}}
HW-only uses only the honest workers for computing $R+\epsilon$ packets, \ie all workers that are not in the set $\Nset_m$, and thus its task completion delay is equal to $T_{\text{HW-only}}=\frac{R+\epsilon}{\sum_{n \notin \Nset_m}1/E[\beta_{n,i}]}$ \cite{KS18}, which can be equivalently written as:
\begin{align}\label{eq:T_trivial}
  T_{\text{HW-only}}=&\frac{R+\epsilon}{\sum_{n \notin \Nset_m}1/E[\beta_{n,i}]}\\
  =&\frac{(R+\epsilon)\sum_{n \in \Nset}1/E[\beta_{n,i}]}{\sum_{n \notin \Nset_m}1/E[\beta_{n,i}]\sum_{n \in \Nset}1/E[\beta_{n,i}]}\\
  =&\frac{\sum_{n \notin \Nset_m} \frac{R+\epsilon}{E[\beta_{n,i}]}}{\sum_{n \notin \Nset_m}1/E[\beta_{n,i}]\sum_{n \in \Nset}1/E[\beta_{n,i}]}+\\
  &\quad \frac{\sum_{n \in \Nset} 1/E[\beta_{n,i}]\sum_{n \in \Nset_m}\frac{R+\epsilon}{E[\beta_n,i]\sum_{n \in \Nset}1/E[\beta_{n,i}]}}{\sum_{n \notin \Nset_m}1/E[\beta_{n,i}]\sum_{n \in \Nset}1/E[\beta_{n,i}]}\\
  =&\frac{R+\epsilon}{\sum_{n \in \Nset}1/E[\beta_{n,i}]}+\frac{\sum_{n \in \Nset_m}\frac{R+\epsilon}{E[\beta_{n,i}]\sum_{n \in \Nset}1/E[\beta_{n,i}]}}{\sum_{n \notin \Nset_m}1/E[\beta_{n,i}]}\\
  =& \frac{R+\epsilon}{\sum_{n \in \Nset}1/E[\beta_{n,i}]}+\frac{\sum_{n \in \Nset_m} z_n}{\sum_{n \notin \Nset_m}1/E[\beta_{n,i}]}
\end{align} 
Using the upper bound on $E[T_{\text{SC}^3}]$ provided in Theorem~\ref{th:T_SC3P}, $T_{\text{HW-only}}-E[T_{\text{SC}^3}]$ can be written as:
\begin{align}\label{eq:T_trivial-T_SC3P_proof}
  &T_{\text{HW-only}}-E[T_{\text{SC}^3}] \ge \\
  &\frac{\sum_{n \in \Nset_m} z_n}{\sum_{n \notin \Nset_m}1/E[\beta_{n,i}]}-\frac{\sum_{n \in \Nset_m}z_n \Big(P+ \rho_c\big(1-P\big)\Big)}{\sum_{n \notin \Nset_m}1/E[\beta_{n,i}]}\\
  & =\frac{\sum_{n \in \Nset_m}z_n(1-\rho_c)\big(1-P\big)}{\sum_{n \notin \Nset_m}1/E[\beta_{n,i}]}\\
  & =\frac{(R+\epsilon) (1-\rho_c) \sum_{n \in \Nset_m}\frac{1-P}{E[\beta_{n,i}]}}{\big(\sum_{n \in \Nset}1/E[\beta_{n,i}]\big) \big(\sum_{n \notin \Nset_m}1/E[\beta_{n,i}]\big)}.
\end{align}
This concludes the proof.

\end{document}